\documentclass[aps,prb,twocolumn,superscriptaddress]{revtex4-2}

\usepackage{graphicx}
\usepackage{amssymb}
\usepackage{amsmath}
\usepackage{amsfonts}
\usepackage{braket,gensymb}
\usepackage{color}
\usepackage[utf8]{inputenc}		
\usepackage[english]{babel}
\usepackage{array}

\begin{document}
\title{
Topological transition in spectrum of skyrmion crystal with uniaxial anisotropy
} 
\author{V.~E. Timofeev}
\email{viktor.timofeev@spbu.ru}
\affiliation{St.Petersburg State University, 7/9 Universitetskaya nab., 199034
St.~Petersburg, Russia} 
\affiliation{NRC ``Kurchatov Institute'', Petersburg Nuclear Physics Institute, Gatchina
188300, Russia}

\author{D.~A. Bedyaev}
\affiliation{St.Petersburg State University, 7/9 Universitetskaya nab., 199034
St.~Petersburg, Russia} 
\affiliation{NRC ``Kurchatov Institute'', Petersburg Nuclear Physics Institute, Gatchina
188300, Russia}

\author{D.~N. Aristov}
\affiliation{NRC ``Kurchatov Institute'', Petersburg Nuclear Physics Institute, Gatchina
188300, Russia}
\affiliation{St.Petersburg State University, 7/9 Universitetskaya nab., 199034
St.~Petersburg, Russia} 

\begin{abstract} 
The band structure of elementary excitations of skyrmion crystal in thin ferromagnetic film with Dzyaloshinskii-Moriya interaction and uniaxial magnetic anisotropy under external magnetic field is studied. 
In the absence of anisotropy there is a topological transition in the spectrum of skyrmion crystal: the gap between breathing and counter-clock-wise modes closes, which is accompanied by changes of Berry curvature sign of these bands. In this work we demonstrate that such topological transition exists in some range of the uniaxial anisotropy values. We present a phase diagram showing that the value of the field of topological transition is higher in the easy-plane domain and lower in the easy-axis domain of anisotropy.
\end{abstract}

\maketitle

{\bf Introduction}. Magnetic skyrmions are whirls of local magnetization in quasi-two-dimensional noncentrosymetric magnets \cite{Nagaosa2013}. Key feature of the skyrmions is their topological charge, an integer number that characterizes the winding degree of each skyrmion's configuration\cite{Belavin1975}. The magnetic skyrmions are an excellent example of objects in condensed matter physics whose nontrivial topology affects physical properties of the system \cite{fert2017magnetic}. Due to skyrmions' presence one observes the topological contribution to electron Hall conductivity\cite{G_bel_2018}, thermal Hall effect for magnons  \cite{PhysRevResearch.4.043085} etc.

In presence of asymmetric Dzyalosinskii-Moriya interaction (DMI) there is a range of external magnetic fields 
where skyrmions are arranged into regular lattices, co-called skyrmion crystals (SkX) \cite{bogdanov1989thermodynamically,BOGDANOV1994255}. The whirling structure of the static skyrmion configuration leads to peculiarities  in gauge potentials defining the magnon equations of motion \cite{schutte2014magnon}, so that the band structure \cite{roldan2016} of elementary excitations in SkX possesses nontrivial topology encoded in Berry curvature and Chern numbers \cite{garst2017collective,Timofeev2022}. It was shown that the low-energy excitations of SkX have a complicated band structure with both topological and trivial bands \cite{Weber_2022,Ghader_2024}.

It was shown that there is a topological transition in the band structure of SkX in a simplest model of two-dimensional ferromagnet with DMI placed into external magnetic field \cite{Diaz2020}. The energy gap between low-energy breathing (Br) and counter-clockwise (CCW) modes at the  Brillouin zone center is closing with the increase of magnetic field. The reopening of the gap with further increase of the field is accompanied by a change in the Chern numbers of these bands \cite{Timofeev_2023c}. In experiments, a tendency toward a decrease of the gap between the Br and CCW modes is observed \cite{onose2012observation,Schwarze2015}, while its reopening was not detected \cite{okamura2013microwave,takagi2021}. 

The present paper is devoted to elucidating the role of uniaxial magnetic anisotropy upon such a topological transition. We briefly discuss the model and methods that we use. We then find the optimal static configuration of SkX, calculate the spectrum and wave functions of elementary excitations and analyze the uniform susceptibility tensor. The crossing of Br and CCW branches is accompanied by the changes in the sign of Berry curvature and we present the phase diagram showing the location of this transition. We present our conclusions in the final paragraphs of the paper. \\

%

{\bf Model}.
We study a model of thin ferromagnetic film with DMI and uniaxial anisotropy under external magnetic field perpendicular to the film, with the following energy density:
\begin{equation}
\mathcal{E} =   \frac{C}{2}  (\nabla \mathbf{S})^2 + 
D(\mathbf{S},\nabla\times\mathbf{S})  - B  S_{z} + A S_z^2\,,
\label{eq:classicalenergy}
\end{equation}
where $C$ is ferromagnetic stiffness, $D$ is DMI constant, $B$ is proportional to the external magnetic field and $A$ is an anisotropy constant. We consider the low temperature case, when the local magnetization is saturated $\mathbf{S}=S\mathbf{n}$ and its orientation doesn't change across the thickness of the film.

In order to simplify the consideration of the model \eqref{eq:classicalenergy}, we introduce the unit system, where length is measured in units of $l=D/C$ and  the energy density in units of $ CS^2 l^{-2} = S^2D^2/C$.   It  is also convenient to combine the last two terms in \eqref{eq:classicalenergy} into a full square, and to count the energy density from the value  $-B^2/4A$. With these agreements the model \eqref{eq:classicalenergy} takes the following form:
\begin{equation}
\mathcal{E} =   \frac{1}{2}  (\nabla \mathbf{n})^2 + 
(\mathbf{n},\nabla\times\mathbf{n})  + a\left( n_{z}-\frac{b}{2a}\right)^2\,. 
\label{eq:classicalenergy2}
\end{equation}
where $b=BC /SD^2$ and $a=AC/D^2$ are dimensionless parameters. The configuration of  local magnetization, corresponding to the ground state of \eqref{eq:classicalenergy2}, depends on these two parameters. We assume that $b>0$ 
while the sign of $a$ may be arbitrary. 
The {\it easy-axis} and {\it easy-plane}  anisotropy correspond to  $a<0$ and  $a>0$, respectively.

The model \eqref{eq:classicalenergy} is well-studied, and its phase diagram at $T=0$ was reported  in the works \cite{PhysRevX.4.031045,Lin2015,G_ng_rd__2016}. It was found that 
for a fixed value of $a$ 
and for a gradual increase  of the magnetic field, $b$, 
the stable configuration of local magnetization $\mathbf{S}$ is described successively by  
 magnetic helix,  SkX, and uniform configuration. This sequence of phases happens in a wide range of moderate  values of $a$, but is not realized for the strong easy-axis anisotropy, $a<-1.2$, with uniform configuration  favorable for all $b$ values, and also for the strong easy-plane anisotropy, $a \gtrsim 2b>1$, where SkX with the square lattice may be realized. In the present work we  focus on the region of moderate anisotropy.\\

{\bf Stereographic projection approach}.
We construct SkX configuration in a framework of a stereographic projection approach:
\begin{equation}
n_x + i \, n_y  = \frac{2f}{1 + f\bar{f}}\,,\quad 
n_z = \frac{1 - f\bar{f}}{1 + f\bar{f}}\,,
\label{eq:stereo}
\end{equation}
where $f$ is a complex valued function depends on spacial coordinates $x$ and $y$. It was shown that stereographic function of multi-skyrmion configuration can be well approximated by a sum of stereographic functions of individual skyrmions \cite{Timofeev2019,timofeev2021}, $f_{multi}(\mathbf{r}) = \sum_jf_1(\mathbf{r}-\mathbf{r}_j)$, with the properly chosen stereographic function for a single skyrmion, $f_1(\mathbf{r})$.

The stereographic function, $f_1(\mathbf{r})$, has a first order pole at the center of skyrmion, and rapidly decreases at distances exceeding the skyrmion radius. This profile function can be found numerically with a high precision,
see\cite{Timofeev2019}, but one can equally well use simple trial functions describing the profile of one skyrmion.

In this work we adapt the well known profile function for individual skyrmion, the so-called $2\pi$ domain wall ansatz:
\begin{equation}
\theta_{1}(r)=\sum\limits_{+-}\arcsin{\left(\tanh{\frac{-r\pm R}{\delta}}\right)} +\pi\,,
\label{eq:cosCMD}
\end{equation}
where $r=\sqrt{x^2+y^2}$, $R$ defines the skyrmion radius and $\delta$ is a width of the domain wall. The magnetization of a single Bloch skyrmion placed at the origin is expressed as follows: $\mathbf{n}(\mathbf{r})=(-y\sin{\theta_{1}(r)}/r,x\sin{\theta_{1}(r)}/r,\cos{\theta_{1}(r)})$. The profile function \eqref{eq:cosCMD} is widely used to describe experimental profiles of individual magnetic skyrmions \cite{Romming_2015}, and the detailed theory of isolated skyrmions is also based on this profile \cite{B_ttner_2018}.

The stereographic function of the Bloch skyrmion with the profile function \eqref{eq:cosCMD} is given by 
\begin{equation}
f_1=\frac{i \cosh{R/\delta}}{\bar{z}}\times\frac{\sqrt{z\bar{z}}}{\sinh{\sqrt{z\bar{z}}/\delta}}\,,
\label{eq:stereoCMD}
\end{equation}
where $z=x+iy$ and $\bar{z}=x-iy$.
Notice, that the skyrmion radius, $R_{sk}$, defined by the condition, $|f_1(\mathbf{r})|=1$, is found from  the equation, 
$\sinh{R_{sk}/\delta}= \cosh{R/\delta} $.

\begin{figure}[t]
\center{\includegraphics[width=0.99\linewidth]{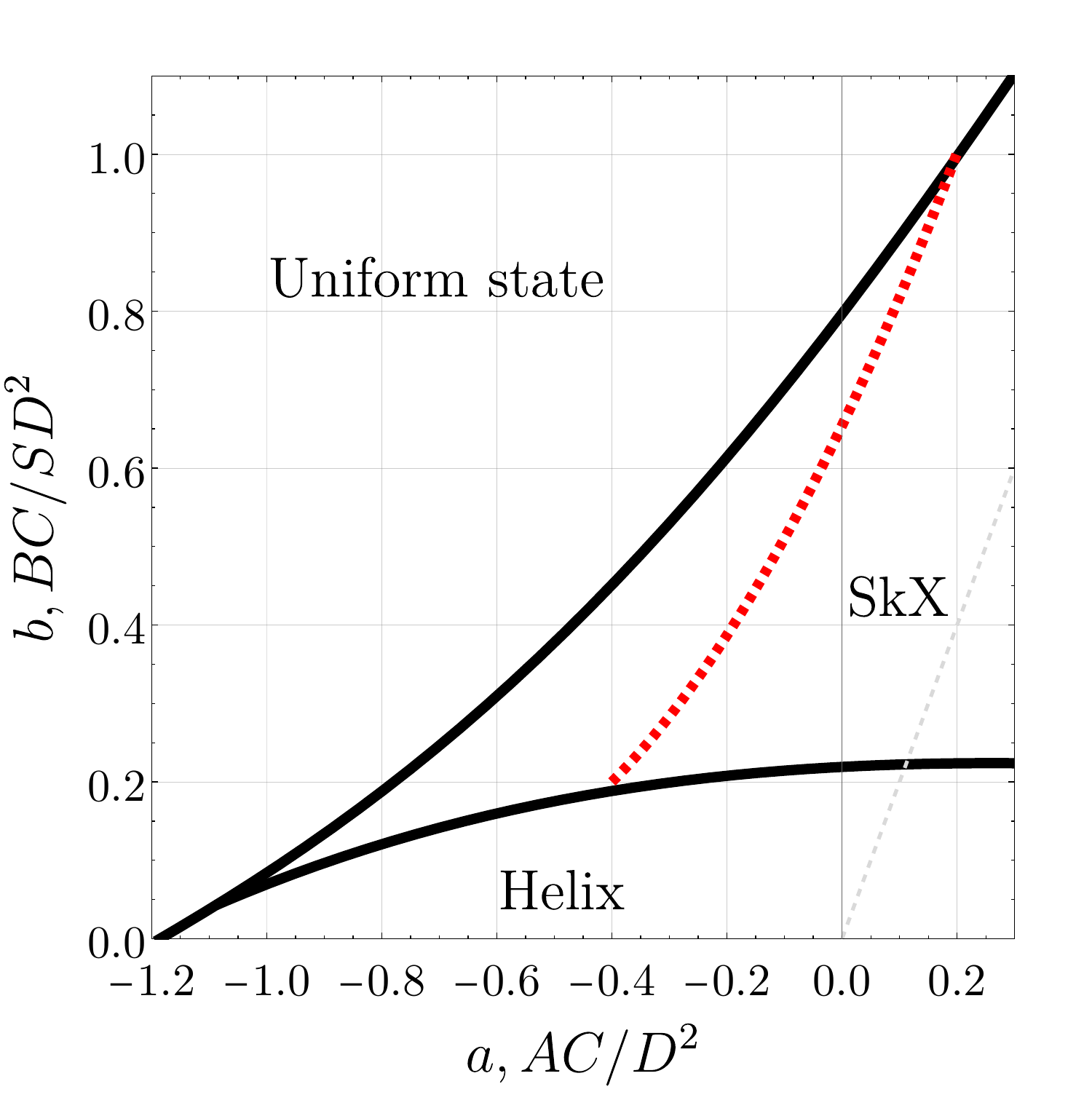}}
\caption{Investigated region of the phase diagram of the model \eqref{eq:classicalenergy2}. The black lines are borders of three  phases: uniform magnetization, SkX and magnetic Helix. The red dashed line marks the topological transition in the spectrum, it corresponds to coincidence of  frequencies of Br and CCW modes at the Brillouin zone center.}
\label{fig:phasediag}
\end{figure}

Using the function 
 \eqref{eq:stereoCMD}, one can construct the stereographic function of SkX in a following way: $f_{SkX}(\mathbf{r}) = \sum_{j}f_1(\mathbf{r}-n_j\mathbf{a}_1-m_j\mathbf{a}_2)$, where the vectors $\mathbf{a}_1 = d\, \mathbf{e}_x$ and $\mathbf{a}_2 =\tfrac12 d\, \mathbf{e}_x+\tfrac{\sqrt{3}}{2}d\,\mathbf{e}_y $ are  primitive translations of triangular lattice. Taking this ansatz for $f_{SkX}(\mathbf{r})$, the energy density of SkX depends on three parameters: $d,R$ and $\delta$. The optimal configuration is  found by minimizing the value $\tfrac1v\int_{cell}d\mathbf{r}\,\mathcal{E}$, where $v$ is the volume of 2D unit cell, for each pair of $a$ and $b$ . The phase diagram obtained with the use of such procedure agrees qualitatively and quantitatively with the results of previous studies \cite{PhysRevX.4.031045,Lin2015,G_ng_rd__2016} for the parameter range of interest, see Fig.\ref{fig:phasediag}.\\

{\bf Semiclassical dynamics of low energy excitations}. The dynamics of local magnetization is described by the well known Landau-Lifshitz (LL) equation, which can be viewed as the Euler-Lagrange equation with a proper form of kinetic term, $\mathcal{T}$, in the Lagrangian, 
$L(f,\bar{f},\partial_tf,\partial_t\bar{f})=\int d\mathbf{r}(\mathcal{T} - \mathcal{E})$. 
In the stereographic projection formalism the kinetic term can be represented  as \cite{metlov13vortex}:
\begin{equation}
\mathcal{T}=  \frac i2 \frac{\bar f \partial_t f - f \partial _t \bar f}
{1+f \bar f}\,.
\label{kinLagrangian}
\end{equation}
where gyromagnetic ratio,  appearing in LL equation, is absorbed into the definition of time, $t$. 

There is a convenient way to study the low energy excitations of topological solitons, see for example \cite{Rajaraman, Rubakov2002}. In our case, we consider infinitesimal time-dependent fluctuation of stereographic function in a following form:
\begin{equation}
 f(\mathbf{r},t) = f_0(\mathbf{r}) + (1 + f_0(\mathbf{r}) \bar{f}_0(\mathbf{r}))\psi(\mathbf{r},t)/\sqrt{2S}\,.\,
\label{eq:fexpan}
\end{equation}
where $f_0(\mathbf{r})$ is a static equilibrium configuration of stereographic projection function. Factor $1/\sqrt{2S}$ helps to ensure the commutation relations for local magnetization components after the procedure of quantization, see \cite{Timofeev_2023a}.  We further assume that $1/\sqrt{2S}$ is a small parameter, that allows us to consider a formal expansion for the Lagrangian density: $\mathcal{L} = \mathcal{L}_0 +   \mathcal{L}_1/\sqrt{2S} +   \mathcal{L}_2/2S + O(S^{-3/2})$, where term $\mathcal{L}_0$ is time independent, $\mathcal{L}_1$ vanishes when  $f_0(\mathbf{r})$ delivers the energy minimum. The term $\mathcal{L}_2$ is a quadratic form in $\psi(t)$ and $\bar{\psi}(t)$ and corresponds to the linear spin-wave theory:

\begin{equation}
\mathcal{L}_2 =\frac{1}{2} 
\begin{pmatrix}
  \bar{\psi},& \psi
\end{pmatrix}
\left(i
\begin{pmatrix}
  \partial_t& 0\\
  0& -\partial_t
\end{pmatrix}
-\hat{\mathcal{H}} \right)
\begin{pmatrix}
  \psi\\
  \bar{\psi}
\end{pmatrix},
\label{Lagr2}
\end{equation}
with the Hamiltonian operator 
$\hat{\mathcal{H}}$ of the form 
\begin{equation}
\hat{\mathcal{H}}=
\begin{pmatrix}
  (-i\nabla + \mathbf{A})^2 + U&
   V\\
  V^*&
   (i\nabla + \mathbf{A})^2 + U
\end{pmatrix} \,,
\label{eq:ham}
\end{equation}
with $\nabla =  \mathbf{e}_x \partial_x + \mathbf{e}_y \partial_y$. 
Here $U$, $V$ and $\mathbf{A} = \mathbf{e}_x A_x + \mathbf{e}_y A_y$ are rather cumbersome  functions of the static function $f_0(\mathbf{r})$ and its gradients, listed in \cite{Timofeev2022}.

The Lagrangian \eqref{Lagr2}  results in 
the Euler-Lagrange equation
\begin{equation}
-i \frac{d}{dt}
\begin{pmatrix}
  \psi\\
  \bar{\psi}
\end{pmatrix}
=\sigma_3\hat{\mathcal{H}}
\begin{pmatrix}
  \psi\\
  \bar{\psi}
\end{pmatrix},
\label{eq:BdGeq}
\end{equation}
and the energies of normal modes, $ \epsilon_n$, are found from
\begin{equation}
\Big( \epsilon_n \, \sigma_3 - \hat{\mathcal{H}} \Big) 
\begin{pmatrix}
  u_n\\
  v_n
\end{pmatrix}  
=  0 \,, 
\label{eq:shr}
\end{equation}
where $\sigma_3$ is the   Pauli matrix. The characteristic equation \eqref{eq:shr}  corresponds to Bogoliubov coefficients, 
$ u_n$,  $v_n$, satisfying orthogonality relation, 
 $ \int d\mathbf{r} \,( u_{n}^{*}  u_{m}  - v_{n}^{*}  v_{m}  )  = \delta_{nm}$, 
see \cite{Timofeev2022} for more details.

For SkX case the normal modes of Eq.\ \eqref{eq:shr} can be found in the Bloch form:
\begin{equation}
 \Psi_{n  \mathbf{k}}(\mathbf{r}) = e^{i\mathbf{k}\mathbf{r}}
\begin{pmatrix}
  u_{n\mathbf{k}}(\mathbf{r})\\
  v_{n\mathbf{k}}(\mathbf{r})
\end{pmatrix}=
e^{i\mathbf{k}\mathbf{r}}\sum\limits_{\mathbf{Q}}e^{i\mathbf{Q}\mathbf{r}}
\begin{pmatrix}
  u_\mathbf{Q}(\mathbf{k})\\
  v_\mathbf{Q}(\mathbf{k})
\end{pmatrix},
\label{eq:Bloch}
\end{equation}
with   $\mathbf{k}$ lying in the first Brillouin zone and the summation is over  reciprocal lattice vectors of SkX, 
$\mathbf{Q}=n\mathbf{b}_1+m\mathbf{b}_2$, with primitive vectors $\mathbf{b}_1 = \tfrac{2\pi}{d}\mathbf{e}_x - \tfrac{2\pi}{\sqrt{3}d}\mathbf{e}_y$ and $\mathbf{b}_1 = \tfrac{4\pi}{\sqrt{3}d}\mathbf{e}_y$. 
We are primarily interested in the topological transition between the low-lying branches of SkX spectrum, which corresponds to gap between CCW and Br modes reopening at $\Gamma$ point, it was observed earlier in the absence of anisotropy, $a=0$, see \cite{Timofeev_2023c}. 
We propose to pinpoint this transition by analyzing the dynamical uniform susceptibility tensor \cite{Timofeev_2023a} and finding a whole set of parameters $a$ and $b$ when the corresponding resonant branches are crossing.\\

{\bf Magnetization quantization and susceptibility tensor}.
The stereographic function \eqref{eq:fexpan} leads to the following form of local magnetization expansion:
\begin{equation}
S_i =S n_i + 
\sqrt{S/2}  ( \bar{F}_i  \psi +  F_i   \bar{\psi})
 + O(1),
\label{Slinear}
\end{equation}
where index $i=1,2,3$, and $\psi$ and $\bar{\psi}$ are expanded in the basis of normal modes of \eqref{eq:BdGeq}. The complex vector $\mathbf{F}$ is expressed in terms of static stereographic function $f_0$:
\begin{equation}
\mathbf{F}=\frac{1}{1 + f_0\bar{f}_0}
\begin{pmatrix}
  1-f_0^2\\
  i(1 + f_0^2)\\
  -2 f_0
\end{pmatrix}\,.
\label{def:F}
\end{equation}
Three vectors, $\mbox{Re }\mathbf{F}$, $\mbox{Im }\mathbf{F}$, and $\mathbf{n}$ form the local orthonormal basis at each point, $\mathbf{r}$, in the plane.

The susceptibility tensor, $\chi_{ij}(\mathbf{k},\omega) =\int dt\,e^{i\omega t} \chi_{ij}(\mathbf{k},t)$,  is the Fourier transform of  the spin retarded Green's function 
\begin{equation}
\chi_{ij}(\mathbf{k},t) = -i \theta(t)\langle[S_i(\mathbf{k},t),S_{j}(-\mathbf{k},0)]\rangle\,.
\label{eq:susctensor}
\end{equation}

As it was mentioned above we are interested in uniform case,  $\mathbf{k}=0$. The susceptibility tensor has  in this case the following form:
\begin{equation}
\begin{aligned}
\chi_{ij}(\omega) &= \chi_\| \hat h_i \hat h_j + \chi_\perp (\delta_{ij} - \hat h_i \hat h_j) +
\chi_{as } \epsilon_{ijk}\hat h_k \,,
\\
 \chi_\| &= 
  \sum\limits_n \frac{ \epsilon_n  S\, |A_{3,n} |^2 }{-(\omega +i\delta)^2+\epsilon_n^2}  \,, 
\\
\chi_\perp &= 
\sum\limits_n \frac{ \epsilon_n S\, | A_{1,n}|^2}{-(\omega +i\delta)^2+\epsilon_n^2}  \,, 
\\
\chi_{as }  &= 
 \sum\limits_n \frac{i\omega S \mbox{ Im}({A}_{1,n}\bar{A}_{2,n} ) }{-(\omega +i\delta)^2+\epsilon_n^2}  \,,
\end{aligned}
\label{chi-comp}
\end{equation}
where vector $\hat h = (0,0,1)$ is normal to the film, the summation index $n$ corresponds to $n$-th spectrum branch, and matrix elements in numerators are defined as :
\begin{equation}
A_{j,n} = \int d\mathbf{r}\,(\bar{F}_ju_n + F_jv_n) \,. 
\label{def:Anj}
\end{equation}

The branches with different symmetries of wave functions contribute differently to various components of susceptibility. 
The breathing mode  manifests itself  to the longitudinal component, $\chi_\|$,  whereas CW and CCW modes contribute to the transverse component, $\chi_\perp$, the difference between the latter two modes is in the sign of the imaginary part of the product, $\mbox{Im}({A}_{1,n}\bar{A}_{2,n})$. Thus, analyzing the resonances in $\chi_\|$, $\chi_\perp$, $\chi_{as }$, we  identify the energy and character of the excitations. 
\\ 

\begin{figure*}[t]
\center{\includegraphics[width=0.99\linewidth]{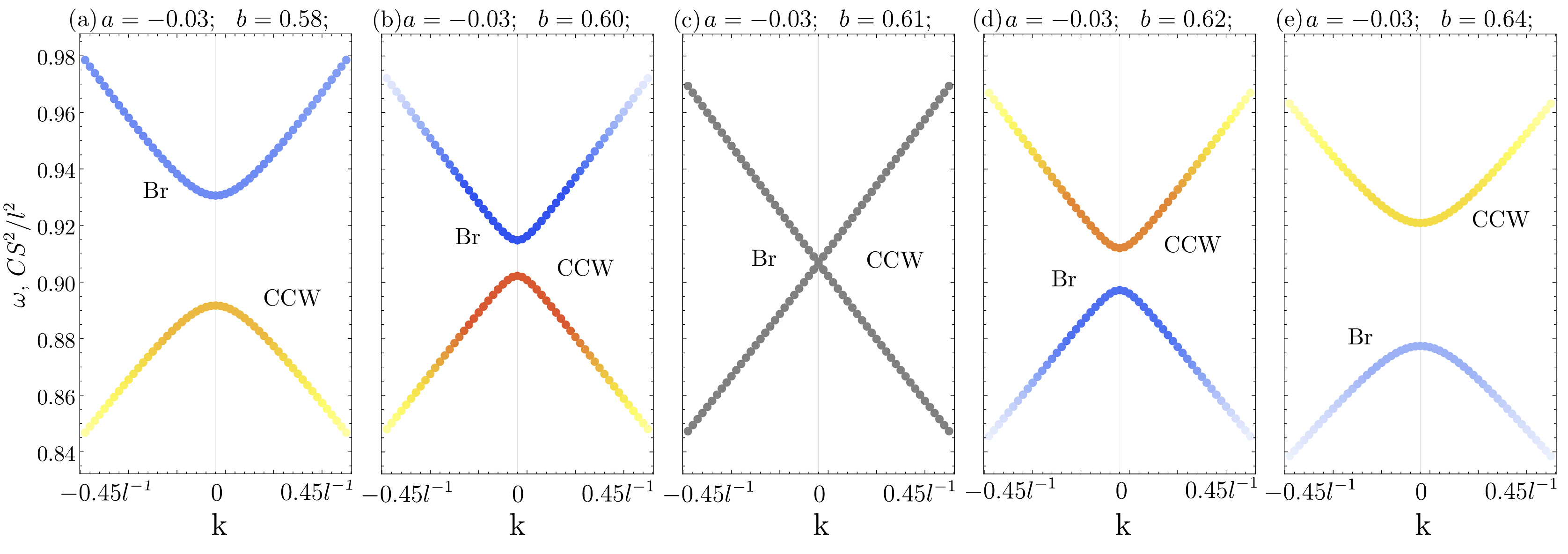}}
\caption{\label{fig:gap}
Panels (a)-(e) demonstrate the evolution of Br and CCW bands for small easy-axis anisotropy, $a=-0.03$, and $b\in[0.58,0.64]$, with the band gap closure at $b=0.61$.   The sign of the Berry curvature for each band is denoted by the color: blue points correspond to negative sign and yellow-red points correspond to positive sign. The similar picture of  the Berry curvature sign change is observed at the whole red dashed line in  Fig.\ \ref{fig:phasediag}.
}
\end{figure*}

{\bf Topology of the band structure}.  We write the Bloch state, Eq.\ \eqref{eq:Bloch}, referring to $n$th band in the form,
$\Psi_{n  \mathbf{k}} = e^{i\mathbf{k}\mathbf{r}}  \mathcal{V}_{n\mathbf{k}} (\mathbf{r})$, 
and assume that it is a smooth function of $\mathbf{k}$. We use the expression for Berry connection 
\begin{equation}
 {{\cal A}}^{\mu}_{n}(\mathbf{k}) = 
-\bra{\mathcal{V}_{n\mathbf{k}}}i \partial_{\mu} \ket{\mathcal{V}_{n\mathbf{k}}}\,, 
\end{equation}
with $\partial_{\mu} = \partial /\partial k_\mu$ ;    the scalar product reads
\[ 
\braket{\mathcal{V}_{n\mathbf{k}} |\mathcal{V}_{n\mathbf{k}'}}
= \sum\limits_{\mathbf{Q}} \left( 
  u^*_{\mathbf{Q}}( \mathbf{k}) 
   u_{\mathbf{Q}}( \mathbf{k}') - 
  v^*_{\mathbf{Q}}( \mathbf{k})
  v_{\mathbf{Q}}( \mathbf{k}')
  \right) \,.
\] 
The Berry curvature is then 
\begin{equation}
\Omega_{n}(\mathbf{k})=\partial_{x}{\cal A}^{y}_{n}(\mathbf{k}) - \partial_{y}{\cal A}^{x}_{n}(\mathbf{k})\,.
\label{eq:berrycurv}
\end{equation}

The integral over the Brillouin zone, $C_n=\tfrac{1}{2\pi}\int_{BZ}\Omega_{n}(\mathbf{k})$, is an integer number for each band and is called Chern number.\\


{\bf Results}. We calculated the energies and wave functions \eqref{eq:shr} of  twenty lowest SkX excitations over the entire range of $a$ and $b$ values shown in Fig.\ref{fig:phasediag}. For each spectral branch, we analyzed the structure of the uniform susceptibility tensor \eqref{eq:susctensor}. In this way, we were able to plot on the phase diagram, Fig.\ \ref{fig:phasediag}, the curve along which the bands corresponding to the Br and CCW excitations cross. This disappearance of the gap  at the center of the Brillouin zone occurs in the range $-0.4\lesssim a \lesssim 0.2$ with the transition field $b_{tt}(a)$ increasing monotonically with increasing $a$.

We confirmed that the gap closing is accompanied by a change in the sign of the Berry curvature \eqref{eq:berrycurv} of the corresponding bands. Figure \ref{fig:gap} shows the evolution of the band structure in the vicinity of the $\Gamma$ point for the anisotropy value $a=-0.03$. Similar behavior of the Berry curvature is observed throughout the entire range $-0.4\lesssim a \lesssim 0.2$ when crossing the field $b_{tt}$.

Generally, the band structure of SkX excitations consists of multiple branches, including  flat bands \cite{roldan2016,garst2017collective} whose energy rapidly changes with the magnetic field. When such flat bands do not overlap with dispersive ones, we can easily calculate their Berry curvature and confirm that the flat bands are topologically trivial \cite{Timofeev2022}. However, if topologically trivial and non-trivial bands intersect, then the calculation of the characteristics of individual bands becomes difficult, and it becomes more appropriate to study the so-called multiband Berry connection and curvature \cite{vanderbilt_2018}, we will discuss this issue in more detail elsewhere.\\



{\bf Discussion}. To the best of our knowledge, the topological transition in the excitation spectrum of SkX has not yet been observed experimentally. It is worth discussing separately why it has not been observed, and how our results relate to experimental findings.

The skyrmion phase in bulk B20-type compounds such as MnSi, Fe$_{1-x}$Ge$_x$, Cu$_2$OSeO$_3$ etc., is observed near the critical temperature, within a narrow range of magnetic fields, $0.4\lesssim H/H_{c2} \lesssim 0.6$\cite{Schwarze2015}. The uniaxial anisotropy in these systems is either absent or negligible, whereas the dipole–dipole interaction acts as easy-plane anisotropy for samples of thin-disk geometry. This latter case corresponds to $a\gtrsim0$ in Fig.\ref{fig:phasediag}, when the field of topological transition is higher than $0.6 H_{c2}$. Therefore, in such systems the gap does not close, although a tendency toward the convergence of resonant frequencies with increasing of magnetic field is regularly observed  \cite{onose2012observation,okamura2013microwave,Schwarze2015,Weber_2022,seki2020propagation,Che_2025}. 
We note that  this tendency holds even in metastable frozen skyrmion phases of Cu$_2$OSeO$_3$ \cite{takagi2021}.

Another class of systems with SkX is given by sandwiches consisting of thin films of ferromagnet and heavy metal Ir/Fe/Co/Pt, Pt/FeCoB etc. In such systems,  DMI is of interfacial nature and the N{\'e}el-type SkX are observed. The easy-axis anisotropy plays a crucial role for skyrmion stabilization in these systems. The experiments on microwave absorption reported both low-frequencies and high-frequencies CCW modes in GHz range, whereas the numerical simulation showed an inverted order of the breathing mode and the CCW mode \cite{Zhang_2017,Satywali2021,Srivastava_2023}, which corresponds to the case $a<-0.4$ in our model, Eq.\ \eqref{eq:classicalenergy2}.

Another system with a significant contribution of easy-axis anisotropy is the polar magnetic semiconductor GaV$_4$S$_8$, where a stable lattice of Néel-type skyrmions forms at low temperatures. It was shown in \cite{Ehlers_2016} that the Br branch lies below the CCW branch in this compound throughout the entire range of magnetic fields.

A change in the ordering of the characteristic resonant frequencies of the Br mode and the CCW mode has been observed in the low-temperature skyrmion (LTS) phase of Cu$_2$OSeO$_3$ \cite{Aqeel2021}. It was shown that cubic anisotropy plays an essential role in stabilizing this phase, while at the same time leading to hybridization of the Br mode with the octupole mode. As a result, the experiment reveals an abrupt change in the resonant frequency of the breathing mode, with $\epsilon_{Br}>\epsilon_{CCW}$  for $H/H_{c2}<0.6$, and $\epsilon_{Br}<\epsilon_{CCW}$  for $H/H_{c2}>0.6$ ; it indirectly points to the topological transition. Changes in the topological properties of each spectral branch were not discussed, and apparently require further investigation.\\

{\bf Conclusions}. We studied the spectra of low-lying excitations of SkX formed in thin ferromagnetic film with DMI, the uniaxial anisotropy and external magnetic field. Within the $2\pi$ domain wall ansatz and stereographic projection approach we constructed optimal SkX configuration for a wide region of the model parameters. 

Calculating the  dispersion and the wave functions for low-energy magnons of SkX, we study the conditions for closing of the gap between experimentally observed Br and CCW modes. We show that the value of magnetic field, when the gap closes, increases with the uniaxial anisotropy shifting from the easy-axis to the easy-plane region in the phase diagram.


Analyzing the wave functions of Br and CCW bands near the center of the Brillouin zone, we show that the gap closing is accompanied by sign changes of the Berry curvature for each band. Thus, we demonstrate that the topological properties of  excitations in skyrmion crystals depend strongly on the model parameters and can be controlled by  the external magnetic field and the value of the anisotropy.

The obtained results are in a good agreement with both the experimental observations and the accompanying calculations and simulations. We hope that our study will stimulate further experimental and theoretical investigations  of SkX dynamics in systems with moderate easy-axis anisotropy, where the topological transition may be directly observed.\\

{\bf Acknowledgments}.
The work was supported by the Russian Science Foundation, Grant No. 24-72-00083. The work of T.V. and D.A. is  partially supported by the Foundation for the Advancement of Theoretical Physics BASIS.

\bibliography{skyrmionbib}

\end{document}